\documentclass[prd,
reprint,
nofootinbib,
 amsmath,amssymb,
 aps,
]{revtex4-1}

\usepackage{graphicx}
\usepackage{dcolumn}
\usepackage{bm}

\newcommand{\dif}{\mathrm{d}}
\newcommand{\e}{\mathrm{e}}
\newcommand{\iu}{\mathrm{i}}
\newcommand{\qcd}{\mathrm{QCD}}
\newcommand{\SU}{\mathrm{SU}}
\newcommand{\SO}{\mathrm{SO}}
\newcommand{\nlsm}{\mathrm{NL\sigma M}}
\newcommand{\trans}{\mathrm{T}}
\newcommand{\tr}{\mathrm{tr}}
\newcommand{\diag}{\mathrm{diag}}

\begin{document}

\preprint{APS/123-QED}

\title{Triggering the QCD phase transition through the Unruh effect: chiral symmetry restoration for uniformly accelerated observers}

\author{Adri\'{a}n Casado-Turri\'on}
\affiliation{%
 Departamento de F\'{i}sica Te\'{o}rica \\
 Universidad Complutense, Madrid, 28040, Spain
}%


\author{Antonio Dobado}
\affiliation{%
 Departamento de F\'{i}sica Te\'{o}rica
}%
\affiliation{%
 Instituto IPARCOS \\
 Universidad Complutense, Madrid, 28040, Spain
}%



\date{\today}


\begin{abstract}
In this work we study the chiral phase transition as observed by an accelerating observer taking into account the Unruh effect.
We use Chiral Perturbation Theory at leading order and the large $N$ limit ($N$ being the number of pions) as an effective description of low-energy QCD, and the Thermalization Theorem to compute the relevant partition function for the accelerating observer. As a result, we obtain that chiral symmetry is restored for uniformly accelerated observers with acceleration $a$ larger than the critical value $a_c=4\pi f_\pi$, with $f_\pi$ being the pion decay constant.
\end{abstract}

\maketitle


\section{Introduction}

It is well known that the formulation of quantum field theory for arbitrary observers in Minkowski spacetime, or in the presence of gravitational fields, is non-trivial mainly due to the possible presence of horizons. The best known examples of this are Hawking radiation \cite{Hawking} and the Unruh effect \cite{Unruh} (see \cite{Crispino:2007eb} for a very complete review). In this work, we shall concentrate on the second one and its connection with chiral symmetry restoration.

Attempting to better understand Hawking radiation, Unruh discovered that the vacuum state of a free quantum field theory in Minkowski spacetime is felt by an uniformly accelerated observer with acceleration $a$ as a thermal ensemble of particles at temperature
\begin{equation} \label{TUnruh}
T=\dfrac{a \hbar}{2\pi c k_B}\simeq 3.97\times 10^{-20}\dfrac{a}{g_\oplus}\text{ K},
\end{equation}
where $g_\oplus\simeq 9.8$ m/s$^2$ is the Earth's mean surface gravity. Therefore, it is thought that the accelerations required to observe the Unruh effect directly are too high to be attained in the laboratory. There exist, however, some proposals for indirect detection. For instance, Bell and Leinaas suggested in 1987 that the Unruh effect may be observed by measuring the polarization of electrons in storage rings \cite{Bell}, but this interesting possibility is still under discussion \cite{Crispino:2007eb}.
 
In spite of this, the Unruh effect is deemed to be highly fundamental. First, it is the result of the interplay between quantum field theory, relativity and statistical mechanics, as can be clearly seen from expression \eqref{TUnruh}, in which the Planck constant $\hbar$, the speed of light $c$ and the Boltzmann constant $k_B$ appear together (in the following we will use natural units with $c=\hbar=k_B=1$). Second, it can be derived in several ways, such as the response to quantum fluctuations of Unruh-DeWitt detectors moving along non-inertial trajectories, canonical quantization \cite{FullingBirrelParkerBoulware} and even axiomatic quantum field theory \cite{Haag:1992hx} in the context of modular theory, where the concept of KMS (Kubo-Martin-Schwinger) states \cite{KMS} plays an essential role \cite{Earman:2011zz}.

Even more importantly, Lee and collaborators showed in 1986   \cite{Lee} that the Unruh effect can be generalized to theories with several interacting fields of arbitrary spin. Moreover, the result does not rely on perturbation theory or any other particular treatment of the interaction. It is thus natural to wonder whether the Unruh effect may be able to trigger non-trivial dynamical effects that appear in finite-temperature interacting field theories in Minkowski spacetime, such as phase transitions. Indeed, it has already been shown that accelerated observers can witness the restoration of spontaneously broken global symmetries in some systems, such as the Nambu-Jona-Lasinio model \cite{Ohsaku:2004rv}, the $\lambda \phi^4$ theory at the one-loop level \cite{Castorina:2012yg} or the Linear Sigma Model (L$\sigma$M) in the large $N$ limit \cite{Dobado}. In fact, Lee's formalism, also known as the Thermalizatiom Theorem, has been successfully employed to study the Brout-Englert-Higgs mechanism (i.e. the restoration of a gauge symmetry) as seen by accelerating observers \cite{Higgs}.

In this work, we shall make use of Lee's formalism to study whether uniform accelerations may lead to a restoration of the spontaneously broken global chiral symmetry of quantum chromodynamics (QCD), the theory of strong interactions in the Standard Model of Particle Physics. With that purpose, we will first review the basic features of this theory in flat spacetime (including its low-energy effective description and the chiral phase transition at finite temperature). After that, we will briefly review the motion of uniformly accelerated observers in Minkowski and the Thermalization Theorem, concluding with the computation of the chiral phase transition in Rindler spacetime, where uniformly accelerated observers live. As expected (for two massless-quark flavors), we will find completely analogous results in both the finite-temperature and uniformly accelerated cases, with a typical second-order (or Landau-Ginzburg) phase transition occurring for accelerated observers with proper acceleration higher than the critical value $a_c=4\pi f_\pi\simeq 1.6$ GeV, where $f_\pi$ is the pion decay constant. There is, however, a significant ---and rather interesting--- difference between both scenarios. In particular, because Rindler spacetime is neither homogeneous nor isotropic (unlike Minkowski), the order parameter of the chiral phase transition (namely, the quark condensate) becomes position-dependent. Thus, it is still possible for a uniformly accelerated observer to perceive a restoration of chiral symmetry even if her acceleration is smaller than the critical one, provided that she lies in the region close to her horizon.

\section{QCD at low energies in Minkowski spacetime}

In this work we will consider a simple version of low-energy QCD consisting in two flavors, $u$ and $d$, forming an $\SU(2)_V$ isospin doublet coupled to an external source $S(x)$. This source is a hermitian color neutral matrix field in flavor space that can be written as $S(x)=s(x)+s^a(x)\,\sigma^a$, with $a=1,2,3$. The partition function, which is a functional on $S$, has an Euclidean path-integral representation given by
\begin{equation} \label{PF1}
Z[S]=\int[\dif G_\mu^a][\dif \bar q][\dif q]\,\exp\bigg(-\int\dif^4 x\,(\mathcal{L}_\qcd+ \bar{q} S q)\bigg),
\end{equation}
where $\mathcal{L}_\qcd$ is the massless QCD Lagrangian,
\begin{equation}
\mathcal{L}_\qcd=-\dfrac{1}{4}G_{\mu\nu}^aG_{\mu\nu}^a+\iu\bar{q}\gamma_\mu D_\mu q,
\end{equation}
with strength tensor $G_{\mu\nu}^a=\partial_\mu G_\nu^a-\partial_\mu G_\nu^a+\iu g f^{abc}G_\mu^b G_\nu^c$ and covariant derivative $D_\mu=\partial_\mu-\iu g T^a G_\mu^a$, and where $T^a$ and $f^{abc}$ respectively denote the generators and structure constants of the $\SU(3)_C$ color gauge group, and thus satisfy $[T^a,T^b]=\iu f^{abc}T^c$.

If $S=0$, the left- and right-chirality parts of the quark fields are independent of each other, and thus the partition function is invariant under global $\SU(2)_L\times\SU(2)_R$ transformations. This symmetry is known as chiral symmetry, and, in general, it is not present if $S\neq 0$. For example, this is the case when we take into account that quarks have small (but still non-vanishing) masses. The mass term for the quarks, $-\bar{q}Mq=-\bar{q}_L M q_R-\bar{q}_R M q_L$, with $M\equiv\diag(m_u,m_d)$ being the quark mass matrix in flavour space, may be introduced in the QCD partition function by setting $S=M$.

The simplest operator in the theory that breaks chiral symmetry explicitly is $\bar{q}q=\bar{u}u+\bar{d}d$. Consequently, its vacuum expectation value (VEV) $\langle\bar{q}q\rangle$, known as the quark condensate, can be considered to be, in some sense, a measure of chiral symmetry breaking. It is not difficult to realize that this VEV may be readily obtained from the dependence of the path integral representation of the QCD partition function on $S$ (in fact, only on $s$) as:
\begin{equation} \label{condensate}
\langle\bar{q}q\rangle=-\dfrac{\delta\ln Z[s]}{\delta s(x)}\bigg|_{s=0}.
\end{equation}
Because Minkowski spacetime is homogeneous and isotropic, the condensate is $x$-independent for inertial observers. However, this is not the case for accelerating observers, as we will see later. We shall also see that, at finite temperature in Minkowski spacetime, the quark condensate may be regarded as the order parameter of the chiral phase transition, being different from zero for temperatures up to a certain critical value $T_c$, and vanishing for $T\geq T_c$.

In addition to its explicit breaking due to the quark masses, it is strongly believed that chiral symmetry is also spontaneously broken into isospin symmetry, i.e. $\SU(2)_L\times\SU(2)_R\rightarrow\SU(2)_{L+R}\equiv\SU(2)_V$, with the three Nambu-Goldstone bosons (NGB) corresponding to the symmetry breaking pattern being the pions. When the small quark masses are taken into account, the pions become massive as well, but are nonetheless much lighter than the rest of the hadrons. Thus, even in this case the low-energy dynamics and the partition function are largely dominated by the pions, whose masses may be considered to be almost negligible.

Since QCD is asymptotically free, its low-energy behaviour becomes non-perturbative. As a result, colored particles (quarks and gluons) cease to be the relevant degrees of freedom of the theory in this regime. This role is taken over by the lightest particles in the spectrum, which are the pions. This reasoning allows for the formulation of a low-energy effective field theory for QCD which possesses all of its symmetries, and whose relevant degrees of freedom are the (pseudo-)NGB associated to the spontaneous breaking of chiral symmetry. The lowest order term of such effective theory is a Non-Linear Sigma Model ($\nlsm$)\footnote{As it is well known  \cite{Weinberg:1978kz}, the full effective field theory for low-energy QCD is Chiral Perturbation Theory (ChPT), which consists of a double expansion in momenta and quark masses at different orders. The leading order in the double expansion corresponds to the $\nlsm$ described above.} based on the coset space $S^3=\SU(2)_L\times\SU(2)_R/\SU(2)_{L+R}=\SU(2)_{L-R}$. Therefore, the low energy representation of the partition function is
\begin{equation} \label{PF2}
Z[S]=\int[\dif\pi^a]\,\exp\bigg(-\int\dif^4 x\,\mathcal{L}_\nlsm\bigg),
\end{equation}
with  $\mathcal{L}_\nlsm$ being the $\nlsm$ Lagrangian density,
\begin{equation} \label{NLSM1}
\mathcal{L}_\nlsm=\dfrac{f_\pi^2}{4}\tr(\partial_\mu U^\dagger\partial_\mu U)-\dfrac{f_\pi^2B}{2}\tr(S(U+U^\dagger)),
\end{equation}
where $f_\pi$ is the pion decay constant, $U(x)\in\SU(2)_{L-R}$ is the NGB field, which may be parametrized in terms of the three pion fields $\pi^a$ as
\begin{equation}
U(x)=\sqrt{1-\dfrac{\pi^a\pi^a}{f_\pi^2}}+\dfrac{\iu\pi^a\sigma^a}{f_\pi},
\end{equation}
and $B$ is another phenomenological parameter, related at tree level with the pion mass $M_\pi$ through $M_\pi^2=2Bm$, with $m=(m_u+m_d)/2$ (this can be easily checked by expanding the $\nlsm$ Lagrangian in the pion fields). For the sake of simplicity, in the following will consider that $m_u=m_d\equiv m$ (the isospin limit). Experimentally, one has $M_\pi\simeq 138$ MeV and $f_\pi\simeq 130$ MeV.

There is a much more useful way of writing the $\nlsm$ Lagrangian, which may be readily generalized to the case in which there are $N$ pions (something which will be crucial later on this work, when we have to consider the large $N$ limit). First, we need to introduce the real quadruplet $\Phi^\trans=(\pi^a,\sigma)$, which belongs to the fundamental representation of $\SO(4)\simeq\SU(2)_L\times\SU(2)_R$. Second, we must notice that the triplet $\pi^a$ belongs to to the fundamental representation of $\SO(3)\simeq\SU(2)_{L+R}$. Thus, the $\nlsm$ coset space may be expressed as $S^3=\SO(4)/\SO(3)$. The same reasoning allows one to obtain the more general result $S^N=\SO(N+1)/\SO(N)$. Thus, from now on, we will let the indices $a,b,\text{etc.}$ run from 1 to $N$, with $N$ being 3 or larger depending on the context.

With this new notation, the non-linear constraint of the model becomes
\begin{equation}
\Phi^\trans\Phi=\pi^a\pi^a+\sigma^2=f_\pi^2,
\end{equation}
which can be solved for $\sigma $, yielding
\begin{equation}
\sigma=\sqrt{f_\pi^2-\pi^a\pi^a}.
\end{equation}
In addition, the $\nlsm$ Lagrangian turns into
\begin{equation} \label{NLSM2}
\mathcal{L}_\nlsm=\dfrac{1}{2}\partial_\mu \Phi^\trans\partial_\mu \Phi-2Bf_\pi s\sigma,
\end{equation}
which may be cast into the alternative form
\begin{equation} \label{NLSM3}
\mathcal{L}_\nlsm=\dfrac{1}{2}\gamma_{ab}\partial_\mu \pi^a\partial_\mu \pi^b-2Bf_\pi^2 s\sqrt{1-\dfrac{\pi^a\pi^a}{f_\pi^2}},
\end{equation}
where
\begin{equation}
\gamma_{ab}=\delta_{ab}+\dfrac{\pi_a\pi_b}{f_\pi^2-\pi^a\pi^a}
\end{equation}
is the coset metric. Notice that \eqref{NLSM3} is entirely written in terms of the pion fields.

The low-energy representation of the QCD partition function may be now reexpressed as
\begin{multline} \label{PF3}
Z[s]=\int[\dif\Phi]\,\delta[\Phi^\trans\Phi-f_\pi^2] \\
\times\exp\bigg(-\int\dif^4 x\,\bigg(\dfrac{1}{2}\partial_\mu\Phi^\trans\partial_\mu\Phi-2Bf_\pi s\sigma\bigg)\bigg),
\end{multline}
where the non-linear constraint is enforced in the path integral through the Dirac delta functional. It is possible to exponentiate the Dirac delta by introducing a non-dynamical Lagrange multiplier field $\lambda(x)$
\begin{multline}
Z[s]=\int[\dif\Phi][\dif\lambda]\,\exp\bigg(-\int\dif^4 x\,\bigg(\dfrac{1}{2}\partial_\mu\Phi^\trans\partial_\mu\Phi \\
+\dfrac{\lambda}{2}(\Phi^\trans\Phi-f_\pi^2)-2Bf_\pi s\sigma\bigg)\bigg).
\end{multline}
The quark condensate computed in this low-energy representation of the QCD partition function using \eqref{condensate} is
\begin{equation} \label{qq}
\langle\bar{q}q\rangle=-2Bf_\pi\langle\sigma\rangle. 
\end{equation}
Furthermore, we also find the lowest-order Gell-Mann, Oakes and Renner relation, $\langle\bar{q}q\rangle=-f_\pi^2 M_\pi^2/m$, which allows us to express the quark condensate in terms of the physical and phenomenological parameters.

\section{The chiral phase transition in the many-pions limit} \label{chiralfiniteT}

In the following, we will set $s$ to $m=M_\pi^2/2B$, so that $Z=Z[m]=Z(M_\pi)$. Therefore,
\begin{multline} \label{PF4}
Z=\int[\dif\Phi][\dif\lambda]\,\exp\bigg(-\int\dif^4 x\,\bigg(\dfrac{1}{2}\partial_\mu\Phi^\trans\partial_\mu\Phi \\
+\dfrac{\lambda}{2}(\Phi^\trans\Phi-f_\pi^2)-M_\pi^2f_\pi\sigma\bigg)\bigg).
\end{multline}
In order to consider the $\nlsm$ at some finite temperature $T=1/\beta$, we must require the various fields to be periodic in Euclidean time with period $\beta$.\footnote{This is because all the fields appearing in \eqref{PF4} are bosonic; fermionic fields at finite temperature should be \textit{anti}-periodic in Euclidean time.} This implies, in turn, that the integral over the Euclidean time must be performed only over the interval $[0,\beta]$.

The previous finite-temperature partition function may be explicitly computed in the large $N$ limit, in which the $\nlsm$ is renormalizable. For this limit to be properly defined, we need to consider $f_\pi^2$ as a quantity of order $N$, i.e. $f_\pi^2\equiv NF^2$, with $F$ being $N$-independent. In such case, the partition function may be written as
\begin{equation} \label{PF4b}
Z=\int[\dif\pi^a][\dif\sigma][\dif\lambda]\,\e^{-\Gamma[\pi^a,\sigma,\lambda]},
\end{equation}
where we have defined the effective action
\begin{multline}
\Gamma[\pi,\sigma,\lambda]=\int\dif^4 x\,\bigg(-\dfrac{1}{2}\pi^a\square\pi^a-\dfrac{1}{2} \sigma\square\sigma \\
+\dfrac{\lambda}{2}(\pi^a\pi^a+\sigma^2- f_\pi^2)-M_\pi^2 f_\pi\sigma\bigg).
\end{multline}
The integral over the $N$ pion fields is Gaussian, and thus it is straightforward to see that it yields
\begin{multline}
\int[\dif\pi^a]\,\exp\bigg(\dfrac{1}{2}\int\dif^4 x\,\pi^a
(-\square+\lambda)\pi^a\bigg) \\
=\exp\bigg(\dfrac{N}{2}\int\dif^4 x\,\ln\dfrac{-\square+\lambda}{-\square}\bigg).
\end{multline}
As a result, we are left with
\begin{equation}
Z=\int[\dif\lambda][\dif\sigma]\,\e^{-\Gamma[\sigma,\lambda]},
\end{equation}
where the effective action in the exponent has reduced to
\begin{multline}
\Gamma[\sigma,\lambda]=\int\dif^4 x\,\bigg(
-\dfrac{1}{2}\sigma\square\sigma+\dfrac{\lambda}{2}(\sigma^2-f_\pi^2) \\
+\dfrac{N}{2}\ln\dfrac{-\square+\lambda}{-\square}-M_\pi^2 f_\pi\sigma\bigg).
\end{multline}
The functional integral above can be computed in the large $N$ limit through the saddle-point approximation, that is, by expanding the fields around some point $(\bar{\sigma},\bar{\lambda})$ in the functional space where the first functional derivatives of $\Gamma[\sigma,\lambda]$ with respect to $\sigma$ and $\lambda$ vanish. Then, by using the steepest descent method one has
\begin{equation} \label{LargeNZ}
Z=\e^{-\Gamma[\bar{\sigma},\bar{\lambda}]}+O(N^{-1/2}), 
\end{equation}
where we have taken into account that $\Gamma[\sigma,\lambda]$ is of order $N$. Therefore, in the large $N$ limit,
\begin{equation}
\bar{\sigma}^2(x)=\langle\sigma\rangle_T^2=\langle\sigma^2\rangle_T.
\end{equation}
with  $\bar{\sigma}$ and $\bar{\lambda}$ being the solutions of:
\begin{eqnarray}
\dfrac{\delta \Gamma}{\delta \sigma(x)} & = & -\square\sigma+\lambda\sigma-M_\pi^2 f_\pi=0, \label{Equ1} \\
\dfrac{\delta \Gamma}{\delta \lambda(x)} &  =  & \dfrac{1}{2}(\sigma^2 -f_\pi^2)+\dfrac{N}{2}G(x,x;\lambda)=0, \label{Equ2}
\end{eqnarray}
where the Euclidean Green function $G(x,x';\lambda)$ satisfies
\begin{equation}
(-\square+\lambda)_x\,G(x,x';\lambda)=\delta^4(x-x').
\end{equation}
In order to solve these equations, it is vital to recall that $\sigma$ and $\lambda$ cannot be position-dependent, because Minkowski spacetime is homogeneous. In the chiral limit ($M_\pi\rightarrow 0$), the first equation simplifies to $\lambda\sigma=0$. Thus, we have two possibilities: a) $\sigma\neq0$, which means that we are on the broken phase with $\langle\bar{q}q\rangle\neq 0$; or b) $\sigma=0$, and thus we must be in the symmetric phase with $\langle\bar{q}q\rangle=0$.

With respect to the second equation, we have, for constant $\lambda$ and thermal boundary conditions,
\begin{equation}
G(x,x;\lambda)= T \sum_{n=-\infty}^{\infty}\int\dfrac{\dif^3\mathbf{k}}{(2\pi)^3}\dfrac{1}{\mathbf{k}^2+\omega_n^2+\lambda}
\end{equation}
where $\omega_n=2 \pi n/\beta$ are the Matsubara frequencies\footnote{Because Euclidean time integrals at finite temperature are performed over the \textit{finite} interval $[0,\beta]$, thus turning temporal Fourier transforms into Fourier series. As a result, the zeroth component of Euclidean momenta are discretized, giving rise to the Matsubara frequencies $\omega_n=2\pi n/\beta$ for bosons (which satisfy periodic boundary conditions) or $\omega_n=2\pi(n+1)/\beta$ for fermions (which satisfy anti-periodic boundary conditions), with $n\in\mathbb{Z}$.} for bosons. For vanishing $\lambda$, this Green function is finite; in fact, it is simply given by
\begin{equation}
G(x,x;0)=T^2/12.
\end{equation}
Thus, the solution of equation \eqref{Equ2} in the chiral limit is
\begin{equation}
\bar{\sigma}=f_\pi^2\bigg(1-\dfrac{T^2}{12 F^2}\bigg),
\end{equation}
where, as we mentioned before, $F^2=f_\pi^2/N$. Introducing the $N$-independent critical temperature
\begin{equation} \label{Tcritical}
T_c^2\equiv 12F^2=\dfrac{12f_\pi^2}{N},
\end{equation}
using equation \eqref{qq}, and requiring the condensate to be a real number, we finally get
\begin{equation}
\dfrac{\langle\bar{q}q\rangle_T}{\langle\bar{q}q\rangle_0}=\begin{cases}
\sqrt{1-\dfrac{T^2}{T_c^2}} & \text{if }0\leq T<T_c, \\
0 & \text{if }T\geq T_c.
\end{cases}
\end{equation}
This result, which corresponds to a typical second-order Ginzburg-Landau phase transition, may also be obtained by performing a diagrammatic computation \cite{Cortes:2016ecy} or by taking the large mass limit for the Higgs-like particle in the corresponding Linear Sigma Model. Furthermore, for $N=3$ and small temperatures compared to $T_c$, we have
\begin{equation}
\dfrac{\langle\bar{q}q\rangle_T}{\langle\bar{q}q\rangle_0}=1-\dfrac{T^2}{8 f_\pi^2}+O\bigg(\dfrac{T^2}{T_c^2}\bigg).
\end{equation}
This result agrees with the one-loop thermal computation in two-flavor Chiral Perturbation Theory  \cite{Gasser:1986vb}.

\section{Rindler spacetime and the Thermalization Theorem} \label{RindlerLee}

From now on, let us denote the usual Cartesian-like coordinates in flat spacetime as $X^\mu=(T,X,Y,Z)$. The Minkowski metric is then
\begin{equation}
\dif s^2=\dif T^2-\dif X^2-\dif X_\perp^2,
\end{equation}
where we have defined $X_\perp\equiv(Y,Z)$ for further simplicity.

Consider an accelerated observer moving with constant proper acceleration $a$ along the $X$ direction of Minkowski spacetime (without loss of generality). Her trajectory is the hyperbola $T^2-X^2=-1/a^2$, whose asymptotes ---the null lines $T=\pm X$--- divide Minkowski spacetime into four quadrants: the regions $\pm X>|T|$, respectively known as the right ($R$) and left ($L$) Rindler wedges; and the regions $\pm T>|X|$, which are the future ($F$) and past ($P$) of the origin, respectively. Notice that one the branches of the hyperbola lies entirely in $R$, while the other one lies in $L$. These two regions are causally disconnected, and thus an accelerated observer in $R$ cannot be affected by any event in $L$, and vice versa. However, both $R$ and $L$ share a common past (region $P$), and so correlations between them may still exist.

Hyperbolic motion is conveniently described using comoving coordinates $x^\mu=(t,x,y,z)$, which can take any real value and are defined by
\begin{equation} \label{Rindlercomov}
\begin{gathered}
T=\dfrac{\e^{ax}}{a}\sinh(at),\hspace{10pt} X=\pm\dfrac{\e^{ax}}{a}\cosh(at), \\
X_\perp=x_\perp,
\end{gathered}
\end{equation}
with the plus-sign choice covering only $R$ and the minus-sign choice covering only $L$. It is important to notice that this implies that the Euclidean Minkowski time $T_E\equiv\iu T$ and $X$ are periodic functions of the Euclidean comoving time $t_E\equiv\iu t$, with period $2\pi/a$. In terms of comoving coordinates, the Minkowski metric is
\begin{equation}
\dif s^2=\e^{2ax}(\dif t^2-\dif x^2)-\dif x_\perp ^2
\end{equation}
in both $R$ and $L$. The worldline with $x=0$ corresponds to the trajectory of constant proper acceleration $a$; in fact, each of the worldlines with constant $x$ is a trajectory with constant proper acceleration $a(x)=a\e^{-ax}$. Hence, it is also useful to define the so-called Rindler coordinates $\rho\equiv 1/a(x)=\e^{ax}/a\in[0,\infty)$ and $\eta\equiv at\in(-\infty,\infty)$, in which the metric becomes
\begin{equation}
\dif s^2=\rho^2\dif\eta^2-\dif\rho^2-\dif x_\perp ^2.
\end{equation}

Let us suppose now that there exists a certain quantum field theory defined over Minkowski spacetime, which describes fields of arbitrary spin and their possible interactions. Consider, without loss of generality, a Rindler observer in $R$ (the results can be immediately generalized to $L$). Because this observer is causally disconnected from all the events in $L$, she will be insensitive to any vacuum fluctuations in that region. As a result, the Minkowski vacuum state of the theory, which is a pure state $|\Omega_M\rangle$ for any inertial observer, becomes a mixed state in the eyes of the accelerated observer, who needs to perform a partial trace over the degrees of freedom in $L$. Lee showed that the corresponding density matrix is
\begin{equation}
\rho_R=\tr_L|\Omega_M\rangle\langle\Omega_M|=\dfrac{\e^{-2\pi H_R/a}}{\tr\,\e^{-2\pi H_R/a}},
\end{equation}
where $H_R$ is the Rindler Hamiltonian (i.e. the generarator of $t$-translations). This density matrix describes a thermal ensemble at temperature $T=a/2\pi$, which is precisely the Unruh temperature originally found for free scalar field theories. However, the applicability of this result is virtually universal: it guarantees that the Minkowski vacuum state will be perceived by a Rindler observer as a thermal state, independently of the field theory considered. This is the reason why it is also known as the Thermalization Theorem.

\section{Chiral symmetry restoration by acceleration}

According to the Thermalization Theorem, it is clear that the restriction of the Minkowski QCD vacuum $|\Omega_\qcd\rangle$ to either Rindler wedge must be a thermal state at temperature $T=a/2\pi$. However, when compared to the Minkowski vacuum states of other quantum field theories, $|\Omega_\qcd\rangle$ is a rather peculiar, highly non-trivial state with a complex structure. In particular, we have already seen that it features the chiral condensate $\langle\bar{q}q\rangle_T$, which is non-vanishing for temperatures below the critical value, signalling the spontaneous breaking of chiral symmetry. It is then natural to wonder whether an accelerated observer perceives a restoration of chiral symmetry if her acceleration $a$ is higher than a certain critical value $a_c$. This issue may be elucidated by functionally quantizing the low-energy effective theory for QCD in the right Rindler wedge (without loss of generality), and then computing the expectation value $\langle\bar{q}q\rangle_a$ on the Minkowski QCD vacuum.

We thus have to repeat the computation detailed in section \ref{chiralfiniteT}, but with two obvious differences. First, we have to define the different fields on the Euclidean version of $R$, instead of the complete Euclidean flat spacetime. Also, we shall use Euclidean Rindler comoving coordinates, $x^\mu=(t_E,x,x_\perp)$, obtained from \eqref{Rindlercomov}. Bearing these facts in mind, the relevant partition function in the chiral limit $M_\pi\rightarrow 0$ is
\begin{equation} \label{PF5}
Z=\int[\dif\pi][\dif\sigma][\dif\lambda]\,\e^{-\Gamma[\pi,\sigma,\lambda]},
\end{equation}
where the effective action in the exponent is
\begin{multline}
\Gamma[\pi,\sigma,\lambda]=\int\dif^4 x\,\sqrt{g}\,\bigg(-\dfrac{1}{2}\pi^a\square\pi^a-\\
-\dfrac{1}{2}\sigma\square\sigma+\dfrac{\lambda}{2}(\pi^2+\sigma^2-f_\pi^2)\bigg),
\end{multline}
with $g$ the determinant of the Euclidean Rindler metric and the various fields satisfying the thermal-like Rindler boundary conditions $\pi^a(t_E=0,\mathbf{x})=\pi^a(t_E=2\pi/a,\mathbf{x})$, $\sigma(t_E=0,\mathbf{x})=\sigma(t_E=2\pi/a,\mathbf{x})$, $\sigma(t_E,|\mathbf{x}|=\infty)=f_\pi$ and $\lambda(t_E=0,\mathbf{x})=\lambda(t_E=2\pi/a,\mathbf{x})$. 
Notice as well that the spatiotemporal integral must be performed on the region $t_E\in[0,2\pi/a]$, $x,y,z\in(-\infty,\infty)$.

We can now proceed formally in the same way as we did in the thermal case in Minkowski spacetime, first integrating out the pions, and then implementing the large $N$ limit through a steepest descent functional integration. By doing this, we find again
\begin{equation}
Z=e^{-\Gamma[\bar{\sigma},\bar{\lambda}]}+O(N^{-1/2}),
\end{equation}
and also
\begin{equation}
\bar{\sigma}^2(x)=\langle\sigma(x)\rangle_a^2=\langle\sigma^2(x)\rangle_a.
\end{equation}
Unlike the situation in Minkowski, we must note though that $\bar{\sigma}$ and $\bar{\lambda}$ shall now depend on $x$, because, as we mentioned before, Rindler space is neither homogeneous nor isotropic. In spite of this added difficulty, both can be chosen again to be the solutions of
\begin{eqnarray}
\dfrac{\delta \Gamma}{\delta \sigma(x)} & = & -\square\sigma+\lambda\sigma=0, \label{Equ1p} \\
\dfrac{\delta \Gamma}{\delta \lambda(x)} &  =  & \dfrac{1}{2}(\sigma^2 -f_\pi^2)+\dfrac{N}{2}G(x,x;\lambda)=0, \label{Equ2p}
\end{eqnarray}
with boundary conditions $\bar{\sigma}=f_\pi$ and $\bar{\lambda}=0$ at $x\rightarrow\infty$, and where the Eculidean Green function now satisfies
\begin{equation}
(-\square+\lambda)_x\,G(x,x';\lambda)=\dfrac{1}{\sqrt{g}}\delta^4(x-x').
\end{equation}

\begin{figure}[h!]
\includegraphics[width=\linewidth]{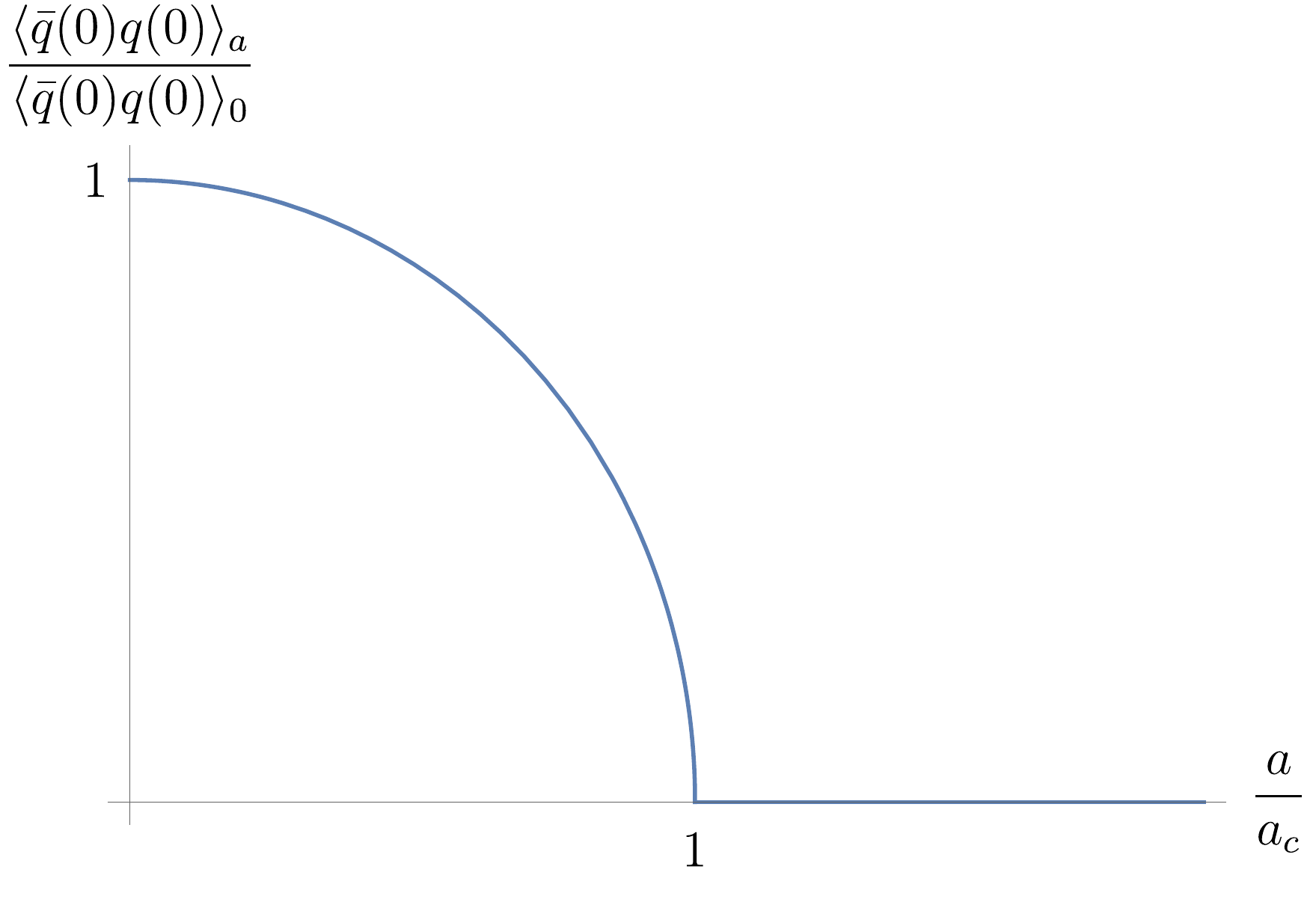}
\centering
\caption{Variation of the quark condensate with $a$, at $x=0$.}
\label{fig:condensate0}
\end{figure}

It is obvious that these equations do not admit constant $\sigma$ and $\lambda$ as solutions, as in the thermal case. Moreover, it is actually very difficult to obtain an exact solution to them. However, they may be solved approximately for $ax<<1$ (i.e. close to the origin of the accelerating frame, $x=0$ or $\rho=1/a$) by exploiting the peculiar properties of Rindler spacetime. With that purpose, we shall follow the steps of \cite{Higgs}, where a similar problem was considered in the context of electroweak symmetry restoration. In the region close to the origin, it is possible to take $\lambda\simeq 0$. Thus, the relevant Euclidean Green function is $G(x,x;0)$, which, as shown in the appendix, is equal to
\begin{equation} \label{Equ2p2}
G(x,x;0)=\int _0^\infty\dif\Omega\,\dfrac{\Omega\pi}{2\rho^2 \tanh(\Omega\pi)}.
\end{equation}
Thus, equation \eqref{Equ2p} transforms into
\begin{equation}
\sigma^2-f_\pi^2+\dfrac{N}{2\pi^3}\int_0^\infty\dif\Omega\,\dfrac{\Omega\pi}{2\rho^2 \tanh(\Omega\pi)}=0.
\end{equation}
Introducing $\omega\equiv a\Omega$ and using $a\rho=\e^{ax}$ we find
\begin{equation}
\sigma^2=f_\pi^2 -\dfrac{N}{4\pi^2}\e^{-2ax}\int _0^\infty\dif\omega\,\omega\,\bigg(1 + \dfrac{2}{e^{2\pi\omega/a}-1}\bigg).
\end{equation}
The first integral on this expression, which is clearly divergent, requires regularization. This can be achieved, for example, by using an ultraviolet cutoff $\Lambda$, leading to an $x$-dependent renormalization of $f_\pi$,
\begin{equation}
f_\pi^2\longmapsto f_\pi^2-\dfrac{N\Lambda^2}{8\pi^2}\e^{-2ax},
\end{equation}
which naturally matches the  $a=0$ limit and is also consistent with the red/blue shift detected by the accelerating observer when receiving a signal emitted at the point $x$. We are thus left with
\begin{equation}
\sigma^2=f_\pi^2-\dfrac{N}{2\pi^2}\e^{-2ax}\int _0^\infty\dif\omega\,\dfrac{\omega}{e^{2\pi\omega/a}-1}.
\end{equation}
Evaluating the remaining integral, and expanding the exponential in $ax<<1$, we obtain
\begin{equation}
\bar{\sigma}^2(x)=f_\pi^2\bigg(1-\dfrac{a^2N}{12 (2\pi)^2 f_\pi^2}(1-2ax)\bigg)+O(a^2x^2).
\end{equation}
Defining the $N$-independent critical acceleration to be
\begin{equation}
a_c^2\equiv\dfrac{48\pi^2 f_\pi^2}{N},
\end{equation}
the previous result may be rewritten as
\begin{equation} \label{linear}
\bar{\sigma}^2(x)=f_\pi^2\bigg(1-\dfrac{a^2}{a_c^2}+2x\dfrac{a^3}{a_c^2}+\ldots\bigg),
\end{equation}
which is also a solution of equation \eqref{Equ1p} at this order. Thus, at the origin of the accelerating frame,
\begin{equation}
\bar{\sigma}^2(0)=f_\pi^2\bigg(1-\dfrac{a^2}{a_c^2}\bigg).
\end{equation}
This implies that the quark condensate (which must be again a real number) depends on $a$ according to
\begin{equation}
\dfrac{\langle\bar{q}(0)q(0)\rangle_a}{\langle\bar{q}(0)q(0)\rangle_0}=\begin{cases}
\sqrt{1-\dfrac{a^2}{a_c^2}} & \text{if }0\leq a<a_c, \\
0 & \text{if }a\geq a_c,
\end{cases}
\end{equation}
as shown in figure \ref{fig:condensate0}. This is exactly the same behaviour we found in the flat-spacetime, thermal case, with the only difference being that, now, $a/a_c$ plays the role of $T/T_c$. Thus, we have again a second order phase transition, which now occurs at the critical acceleration $a_c$. Notice that the Unruh temperature $a_c/2\pi$ corresponding to this critical acceleration is precisely the critical temperature $T_c$ we found in \eqref{Tcritical}.

\begin{figure}[h!]
\includegraphics[width=\linewidth]{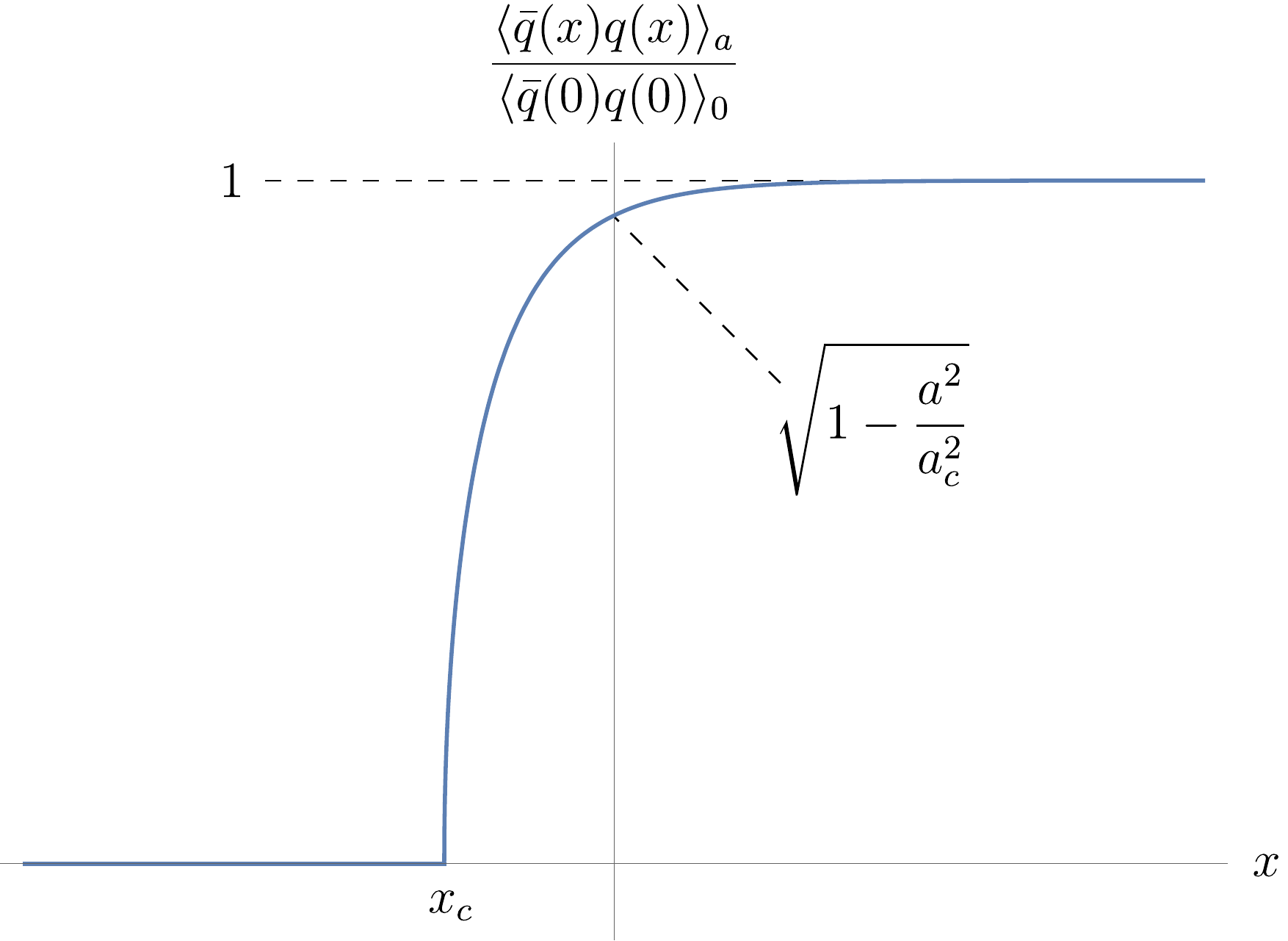}
\centering
\caption{Variation of the quark condensate with $x$, for a given value of $a$ with $0< a<a_c$.}
\label{fig:condensatex}
\end{figure}

The previous result, although very satisfactory, is only valid at $x=0$. In order to study the restoration of chiral symmetry for accelerated observers at points different from the origin, we can make use of the $x=0$ result we have just obtained and some basic properties of Rindler spacetime. Let us consider two accelerated observers with Rindler coordinates $\rho=1/a$ and $\rho'=1/a'$. The results obtained by both observers are exactly the same, but exchanging $a$ and $a'$. From the point of view of the first observer, the second one is located at some point with $x$ coordinate given by
\begin{equation}
\rho'=\dfrac{1}{a'}=\dfrac{\e^{a x}}{a},
\end{equation}
i.e. the acceleration of the second observer is $a'=a\e^{-ax}$. Then, it is immediate to find that the position-dependent value of the condensate is given by
\begin{equation} \label{result}
\dfrac{\langle\bar q(x) q(x)\rangle_a}{\langle\bar q(0) q(0)\rangle_0}=\sqrt{1-\dfrac{a^2}{a_c^2}\e^{-2ax}}.
\end{equation}
Notice that we could have also obtained this result by retaining the full exponential in equation \eqref{linear} and defining the $x$-dependent acceleration $a(x)\equiv a\e^{-ax}$ just as we did in section \ref{RindlerLee}. Therefore, for a Rindler observer with proper acceleration $a\in(0,a_c)$, the condensate is a function of $x$ ranging from $\langle\bar{q}(0)q(0)\rangle$ at infinity to zero at the critical value
\begin{equation}
x_c=\dfrac{1}{a}\ln\bigg(\dfrac{a}{a_c}\bigg)<0,
\end{equation}
as shown in figure \ref{fig:condensatex}.

\begin{figure}[tb!]
\includegraphics[width=0.9\linewidth]{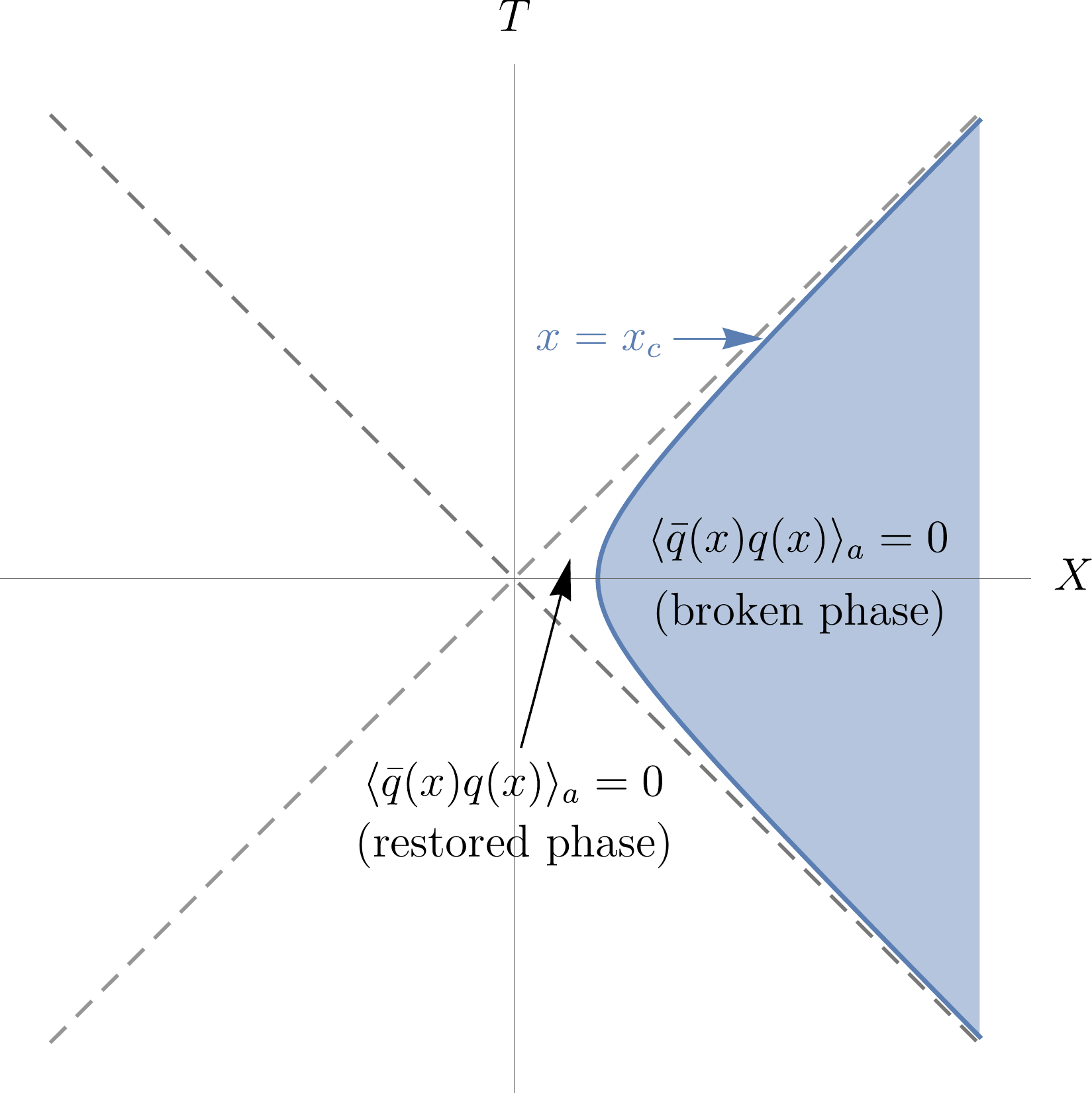}
\centering
\caption{Quark condensate  for accelerating observers. The blue hyperbola marks the $(x=x_c,x_\perp)$ surface at which the phase transition occurs, separating the broken phase (in blue) and the symmetric phase (not colored). The horizon is represented by the two dashed lines.}
\label{fig:landscape}
\end{figure}

This means that the chiral phase transition takes place on the $(x= x_c,x_\perp)$ surface. The symmetry is also restored on the region close to the horizon $x< x_c$, where $\langle\bar{q}(x)q(x)\rangle=0$ (see figure \ref{fig:landscape}). Thus, the boundary between the broken and restored phases is completely determined by $a_c$, which  depends only on QCD parameters.

Another important point is that, introducing the Unruh-like critical temperature $T_c=a_c/2\pi$, and also
\begin{equation} \label{Temperaturex}
T(x)\equiv\dfrac{a}{2\pi}\e^{-ax},
\end{equation}
the condensate in the comoving frame is given by
\begin{equation}
\dfrac{\langle\bar q(x) q(x)\rangle_a}{\langle\bar q(0) q(0)\rangle_0}=\sqrt{1-\dfrac{T^2(x)}{T_c^2}}.
\end{equation}
In other words, this implies that, for points different to the origin, what the condensate feels is equivalent \cite{Candelas:1977zza} to a thermal bath with an $x$-dependent temperature given by \eqref{Temperaturex}, which diverges at the horizon and goes to zero at infinity. This temperature is compatible with the Tolman and Ehrenfest rule \cite{Tolman:1930ona} for thermal equilibrium in static space-times, since
\begin{equation} 
T(x)\sqrt{g_{00}}=T(0)\,\e^{-ax}\,\e^{ax}=T(0)=\dfrac{a}{2\pi}
\end{equation}
is an $x$-independent constant, as required by the rule.

Therefore the chiral condensate is position dependent for Rindler observers. It would be interesting to see which is the case for more complicated spaces such as, for example, the time-dependent, spherically symmetric generalization of ordinary Rindler space considered in \cite{Balasubramanian:2013rqa}. However, that study is very involved and beyond the scope of the present work.

\section{Conclusions}

The powerful Thermalization Theorem formalism developed by Lee has allowed us to demonstrate the ability of the Unruh effect to produce non-trivial dynamical effects. In particular, we have been able to study the restoration of chiral symmetry by acceleration. We have shown that analogous results are found in both the thermal and Rindler cases, with a typical second-order phase transition occurring for those accelerated observers whose accelerations are higher than the critical value $a_c=4\pi f_\pi\simeq 1.6$ GeV for $N=3$ pions. We have also shown that chiral symmetry is also restored close to the horizon of any accelerated observer. The behavior obtained for the spacetime-dependent temperature felt by the condensate is compatible with the standard requirements for thermodynamic equilibrium in static spacetimes, of which Rindler space is a particular example.

It is hoped that delving deeper into our results will lead to some interesting applications. For instance, the Unruh effect has been proposed as a possible thermalization mechanism in ultra-relativistic heavy-ion collisions \cite{Kharzeev}, and, consequently, future analyses based on this work may help to shed light on the issue. Even more in  \cite{Castorina:2007eb}
the Hawking-Unruh radiation, as applied to the specific QCD case, could provide a viable account for the thermal behavior observed in multihadron production even in $e^+e^-$ production.

In addition it is also possible that, at least in principle, our results might be relevant for black holes and as well as in some cosmological scenarios, since local Rindler coordinates can always be defined close to their horizons  \cite{Padmanabhan}.

 Finally it is interesting to realize that  phase transitions, as the one considered in this work,  can also  be triggered by spacetime curvature, as it is for example the case of the scalar field in a static Einstein Universe considered in \cite{Denardo:1981zq}. In principle, the effect of symmetry restoration by acceleration considered here has a different nature but it is clear that they must be connected in some way due to the Equivalence Principle. However the Einstein Universe is horizon free whilst accelerating observer horizons seems to play a decisive role in the Unruh effect. Obviously this is a very interesting problem which deserves further study for clarification.   

\section{Acknowledgments}
A. D. thanks Luis \'Alvarez-Gaum\'e  and C. Pajares for very interesting comments and discussions and the CERN Theory Unit, where part of this work was done, for its hospitality. Work supported by the Spanish grant FPA2016-75654-C2-1-P.

\appendix*
\section{The Euclidean Green function $G(x,x;\lambda)$}

The Euclidean Green function $G(x,x';\lambda)$ is defined by
\begin{equation}
(-\square+\lambda)_x G(x,x';\lambda)=\dfrac{1}{\sqrt{g}}\delta^4(x-x'),
\end{equation}
for constant $\lambda$ and the appropriate periodic boundary conditions. Our goal is to solve this equation to find $G(x,x;\lambda)$. The first step is to notice that the Euclidean d'Alembertian in Rindler coordinates is given by
\begin{equation}
\square=\dfrac{1}{\rho^2}\partial_\eta^2+\partial_\rho^2+\dfrac{1}{\rho}\partial_\rho+\nabla_\perp^2.
\end{equation}
We can now introduce the partial Fourier transform
\begin{multline}
G(\rho,\rho',x_\bot-x'_\perp,\eta-\eta';\lambda)=\sum_{n=-\infty}^\infty\e^{\iu n(\eta-\eta')} \\
\times\int\dfrac{\dif^2 k_\perp}{(2\pi)^2}\,\e^{\iu k_\perp(x_\bot-x'_\perp)}\,\tilde{G}(\rho,\rho',k_\perp,n;\lambda)
\end{multline}
which satisfies
\begin{equation}
(\rho^2\partial_\rho^2+\rho\partial_\rho-\alpha^2\rho^2-n^2)\,\tilde{G}=-\rho\delta(\rho-\rho'),
\end{equation}
where $\alpha^2\equiv k^2_\perp+\lambda$. The solution may be written as
\begin{equation}
\tilde{G}(\rho,\rho',k_\bot,n;\lambda)=\int_0^{\infty}\dif\Omega\,\dfrac{\Psi_\Omega(\rho)\Psi_\Omega(\rho')}{\Omega^2+n^2},
\end{equation}
where $\Psi_\Omega(\rho)$ can be obtained from the modified Bessel functions of the second kind with imaginary parameter:
\begin{equation}
\Psi_\Omega(\rho)=\dfrac{1}{\pi}\sqrt{2\Omega\sinh(\Omega\pi)}K_{\iu\Omega}(\alpha\rho).
\end{equation}
By using well known properties of these functions and
\begin{equation}
\sum_{n=-\infty}^{\infty}\dfrac{1}{\Omega^2+n^2}=\dfrac{\pi}{\Omega}\dfrac{1}{\tanh(\Omega
\pi)}
\end{equation}
it is possible to find
\begin{multline}
G(x,x;\lambda)=\dfrac{1}{2\pi^3}\int_0^\infty\dif\Omega\,\cosh(\Omega\pi) \\
\times\int_0^\infty\dif|k_\perp|\,|k_\perp|\,K_{\iu\Omega}^2(\alpha\rho).
\end{multline}
This last integral may be readily solved for $\lambda=0$, yielding \eqref{Equ2p2}.

\end{document}